\documentclass[twocolumn, showpacs,preprintnumbers,amsmath,amssymb]{revtex4}

\usepackage[T1]{fontenc}
\usepackage{ae,aecompl}
\usepackage{color}

\usepackage{graphicx}	
\usepackage{amsmath}	

\usepackage{dcolumn}
\usepackage{bm}

\newcommand{\CUTpcut}[1]{}

\newcommand{\Kmat}{\mbox{\boldmath $\mathcal{K}$}}

\newcommand{\V}{\mbox{\boldmath $V$}}

\newcommand{\Omat}{{\bf O}}

\newcommand{\MV}{\mathcal{V}}

\begin{document}
\preprint{APS/123-QED}


\title[Population of ground and lowest excited states of Sulfur via DR]{Population of ground and lowest excited states of Sulfur via the dissociative recombination of SH$^+$ in the diffuse interstellar medium}
\author{J.~Boffelli,$^{1}$}
\author{F.~Gauchet,$^{1}$}
\author{D.~O.~Kashinski,$^{2}$}
\author{D.~Talbi,$^{3}$}\email[]{dahbia.talbi@umontpellier.fr}
\author{A.~P.~Hickman,$^{4}$}
\author{K.~Chakrabarti,$^{5}$}
\author{E.~Bron,$^{6}$}
\author{A.~Orb\'an,$^{7}$} 
\author{J.~Zs.~Mezei,$^{1,7}$}\email[]{mezei.zsolt@atomki.hu}
\author{I.~F.~Schneider$^{1,8}$}
\affiliation{$^1$Laboratoire Ondes et Milieux Complexes, LOMC-UMR6294, Universit{\'{e}} Le Havre Normandie, 76058 Le Havre, France}
\affiliation{$^2$Department of Physics and Nuclear Engineering, United States Military Academy, West Point, 10996 New York, USA}
\affiliation{$^3$Laboratoire Univers et Particules de Montpellier, Université de Montpellier--CNRS, 34095 Montpellier, France}
\affiliation{$^4$Department of Physics, Lehigh University, Bethlehem, 18015 Pennsylvania, USA}
\affiliation{$^5$Department of Mathematics, Scottish Church College, 700006 Kolkata, India}
\affiliation{$^6$Laboratoire d’Etudes du Rayonnement et de la Matière en Astrophysique et Atmosphères, LERMA-UMR8112, Observatoire de Paris, \\\,\,\,\,PSL Research University, CNRS, Sorbonne Universit\'es, 92190 Meudon, France}
\affiliation{$^7$Institute for Nuclear Research (ATOMKI), H-4001 Debrecen, Hungary}
\affiliation{$^8$Laboratoire Aim\'{e}-Cotton, LAC-UMR9188, Universit\'e Paris-Saclay, 91405 Orsay, France}
\date{\today}



\begin{abstract}
Our previous study on dissociative recombination of ground state SH$^+$ into $^2\Pi$ states of SH is extended by taking into account the contribution of 
$^4\Pi$ states recently explored by quantum chemistry methods. Multichannel quantum defect theory is employed for the computation of cross sections and rate coefficients  for dissociative recombination, but also for vibrational excitation. Furthermore, we produce the atomic yields resulting from recombination, quantifying the generation of sulfur atoms in their ground (\mbox{$^3$P}) and lowest excited (\mbox{$^1$D}) states respectively. 
\end{abstract}



\maketitle


\section{Introduction}

SH$^+$ (sulfoniumylidene or sulfanylium) is ubiquitous in the Interstellar Medium (ISM). It has been detected in emission in W3 IRS 5, a region of high-mass star formation~\cite{Benz2010}, in absorption in the diffuse interstellar medium, towards various distant star-forming regions~\cite{Menten2011,Godard2012} and in emission in the Orion Bar, a typical high UV-illumination warm and dense photon-dominated region (PDR)~\cite{ZNagy2013}.

Thanks to this ubiquity, the SH$^+$ ion provides unique physical and chemical constraints on models that are applied to the environments where it is observed. SH$^+$ is suggested to form from the reaction of atomic S$^+$ with H$_2$. However this reaction is highly endothermic by $0.86$ eV (9860 K). To overcome this endothermicity in the diffuse ISM, turbulent dissipation, shocks, or shears are invoked. As a consequence, SH$^+$ is suggested as an important probe of turbulence~\cite{Godard2012,Godard2014} of the diffuse ISM. In photon dominated region where H$_2$ is vibrationally excited, it is suggested that this excess of vibrational energy allows the formation of SH$^+$ from S$^+$ and H$_2$~\cite{Zanchet2013}, providing information on the temperature of such environments.  However, for SH$^+$ to be used as a tracer of the physical conditions of the media where it is observed, its chemistry needs to be known in detail and in particular its destruction mechanisms.  SH$^+$ does not react quickly with H$_2$~\cite{Millar1986} the most abundant interstellar molecule due to the endothermicity of the reaction, and therefore it is not severely depleted through this reaction. It is therefore considered that dissociative recombination (DR) of this ion with electrons (eq. (\ref{eq:DR})) is an efficient destruction pathway, and a rate constant of $10^{-6}$ cm$^3\cdot$s$^{-1}$ is assigned to this reaction in the KIDA and UMIST astrochemical data bases~\cite{kida,umist}. To validate this assumption we have started few years ago an extensive and accurate study of the DR of SH$^+$ using methods of theoretical chemistry and quantum dynamics~\cite{kashinski2017}. The large variety of theoretical studies undertaken on SH$^+$  as well as on SH  can be found in our previous papers~\cite{kashinski2017,kashinski2021}.

In the present article, we aim to increase the accuracy of our previous computations  \cite{kashinski2017} on the dissociative recombination of the ground-state SH$^{+}$: 
\begin{equation}\label{eq:DR}
 \mathrm{SH}^+(v_i^+)  + e^-(\varepsilon) \rightarrow   \mathrm{S} + \mathrm{H},
\end{equation}
 taking into account the new molecular structure data characterizing the states of $^4\Pi$ symmetry of SH~\cite{kashinski2021}. In addition, we calculate atomic yields produced via the DR process, resulting in the formation of the atomic sulphur in its ground state - $^3$P - and in its lowest excited state - $^1$D - and we produce cross sections and thermal rate coefficients for the vibrational excitation (VE) of the ion:
\begin{equation}\label{eq:VE}
 \mathrm{SH}^+(v_i^+)  + e^-(\varepsilon) \rightarrow   \mathrm{SH}^+(v_f^+)  + e^-(\varepsilon^\prime).
\end{equation}
 in competition with its DR. 
\noindent Here $v_i^+/v_f^+$ stand for the initial/final vibrational quantum numbers of the cation and $\varepsilon/\varepsilon^\prime$ the kinetic energy of the incident/scattered electrons.

The paper is organized as follows: Section \ref{sec:theo} describes the MQDT method and the molecular data set used in the dynamical calculations. Section \ref{sec:results} presents the obtained dissociative recombination and vibrational excitation cross sections and thermal rate constants, as well as the estimated atomic yields and branching ratios. Section \ref{sec:concl} contains the concluding remarks.

\section{Theoretical background}\label{sec:theo}

The MQDT approach has been proven to be a powerful method for the evaluation of the cross sections of the DR \cite{giusti80,seaton1983,jungen96} and competitive processes like ro-vibrational and dissociative excitations.
It was applied with great success to several diatomic systems like H$_2^+$ and its isotopologues \cite{giusti83,ifs-a09,takagi93,tanabe95,ifs-a18,amitay99,epee2015,stac501}, O$_2^+$ \cite{ggs91,guberman-dr99}, NO$^+$ \cite{sn90,ifs-a22,ifs-a36,Ngassam1,motapon06b}, CO$^+$ \cite{mjzs2015}, N$_2^+$ \cite{jtx,N2+2021}, BF$^+$ \cite{mjzs2016}, triatomics like H$_{3}^{+}$ \cite{ifs-a26,kokoou01,kokoou03}, BF$^+_2$ \cite{Kokoouline_2018} and complex organic molecules \cite{acs2019}. 

\subsection{MQDT method} 

The theoretical summary given below is limited to the account of the {\it vibrational} structure and couplings of SH$^+$ and SH, the rotational effects being neglected, and focuses on DR - equation~(\ref{eq:DR}). However, the reader should keep in mind that the other competitive processes (eq.~(\ref{eq:VE})) - such as {\it superelastic collision or vibrational de-excitation} (SEC or VdE) ($v^+_i>v^+_f$), and {\it inelastic collision or vibrational excitation} (IC or VE) ($v^+_i<v^+_f$) - occur simultaneously and display quite similar features.

The DR results from the quantum interference between the \textit{direct} mechanism, involving the doubly excited resonant states SH$^{**}$ and the \textit{indirect} one, occurring via Rydberg singly-excited predissociating states SH$^{*}$.

 The MQDT treatment of DR and VE requires the a priori knowledge of the potential energy curves (PECs) of the ion (ground and excited states), the relevant doubly excited dissociative states of the neutral molecule, and the series of 
 mono-excited 
 Rydberg states - of ground and excited ionic core - conveniently represented by their quantum defects. The driving forces of the DR and VE are the electronic couplings connecting the ionization continuum to the dissociation one.
 
A detailed description of our theoretical approach was given in \cite{jtx,acs2019}, its main steps are the following:

 \textit{i) Building of the interaction matrix} $\boldsymbol{\mathcal{V}}$: For a given symmetry $\Lambda$ of the neutral (electron + ion ({\it core})) system and
assuming that only one partial wave of the incident electron contributes
to the relevant interactions, the geometry($R$)-dependent electronic coupling of an ionization channel relying on the electronic-core state $c_\beta$ ($\beta = 1$ for the ground state (core 1) and  $\beta = 2$ for the excited state (core 2)) with the dissociation channel $d$ \mbox{can be written:}
\begin{align}\label{eq:elcoup}
&&\MV^{(e)\Lambda}_{d,c_\beta}(R) =
\langle\Phi_{d}|H_{el}|\Phi^{el,c_\beta}\rangle,\: \beta = 1, 2
,
\end{align}
where
$\MV^{(e)\Lambda}_{d,c_\beta}(R)$ is assumed to be independent of the energy of the external electron. 

The integration is performed over the electronic coordinates of the neutral system. Here $H_{el}$ denotes the electronic Hamiltonian, $\Phi_{d}$ is the electronic wave function of the dissociative state and $\Phi^{el,c_\beta}$ is the wave function describing the molecular system ``Rydberg electron + ion in its $c_\beta$ electronic state''.

Similarly, the electronic coupling between the two ionization continua is:
\begin{align}\label{eq:elcoupCores}
&&\MV^{(e)\Lambda}_{c_1,c_2}(R) =
\langle\Phi^{el,c_1}|H_{el}|\Phi^{el,c_2}\rangle.
\end{align}

Integrating these couplings over the internuclear distance leads to the non-vanishing elements of the interaction matrix $\boldsymbol{\mathcal{V}}(E)$: 

\begin{align}
&&V_{d,v_{c_\beta}}^{\Lambda}(E)  =  \langle F_{d}(E)|\MV^{(e)\Lambda}_{d,c_\beta}(R)|\chi_{v_{c_\beta}}
	\rangle,   \: \beta = 1, 2.     \label{eq:Vdv}
\end{align}
\begin{align}
&&V_{v_{c_1},v_{c_2}}^{\Lambda}  = \langle \chi_{v_{c_1}}|\MV^{(e)\Lambda}_{c_1,c_2}(R)|\chi_{v_{c_2}}
	\rangle.     \label{eq:Vvv}
\end{align}
Here $\chi_{v_{c_\beta}}$ ($\beta$ = 1, 2) is the vibrational wave function associated with an ionization channel relying on the core $c_\beta$, $F_{d}$ is the regular radial wave function of the dissociative state $d$ and E is the total energy of the molecular system. This interaction is effective at small electron-ion and internuclear distances, in the reaction zone.

\textit{ii) Computation of the reaction matrix} $\boldsymbol{\mathcal{K}}$: Using the second-order perturbative solution of the Lippmann-Schwinger integral equation \cite{Ngassam2,florescu2003,motapon06b}, written in operator form as:
\begin{align}\label{eq:solveK}
&&\boldsymbol{\mathcal{K}}= \boldsymbol{\mathcal{V}} + \boldsymbol{\mathcal{V}}{\frac{1}{E-\boldsymbol{H_0}}}\boldsymbol{\mathcal{V}},
\end{align}
where $\boldsymbol{H_0}$ is the Hamiltonian of the molecular system under study, with the inter-channel interactions neglected.\\
The reaction matrix $\Kmat$ in block form is:
\begin{align}\label{Kmat2}
&&
\Kmat = \left( \begin{array}{ccc}
\Kmat_{\bar d\bar d} & \Kmat_{{\bar d}{\bar v}_{c_1}} & \Kmat_{\bar d {\bar v}_{c_2}}\\
\Kmat_{{\bar v}_{c_1} \bar d} & \Kmat_{{\bar v}_{c_1} {\bar v}_{c_1}} & \Kmat_{{\bar v}_{c_1} {\bar v}_{c_2}}\\
\Kmat_{{\bar v}_{c_2} \bar d} & \Kmat_{{\bar v}_{c_2} {\bar v}_{c_1}} & \Kmat_{{\bar v}_{c_2} {\bar v}_{c_2}}\\
\end{array} \right),
\end{align}
where the collective indices $\bar d$, $\bar v_{c_1}$, $\bar v_{c_2}$  span the ensembles of all individual indices $d$, $v_{c_1}$ and $v_{c_2}$, which respectively label dissociation channels, ionization channels built on core 1 and ionization channels built on core 2.

An extensive and rigorous derivation of the structure of each block of the $\Kmat$-matrix in second order for a multi-core case was provided in our earlier work \cite{kc2013a}. For SH$^+$, with two attractive ion cores, a natural application of our earlier work leads to the following form of the $\Kmat$-matrix in second order:
\begin{align}\label{Kmat2*}
&&\Kmat = \left( \begin{array}{ccc}
\Omat & \V_{\bar d \bar v_{c_1}} & \V_{\bar d \bar v_{c_2}} \\
\V_{\bar v_{c_1} \bar d} & \Kmat_{\bar v_{c_1} \bar v_{c_1}}^{(2)} & \V_{\bar v_{c_1} \bar v_{c_2}}\\
\V_{\bar v_{c_2} \bar d} & \V_{\bar v_{c_2} \bar v_{c_1}} & \Kmat_{\bar v_{c_2} \bar v_{c_2}}^{(2)}\\
\end{array} \right),
\end{align}
where the elements of the diagonal blocks of $\Kmat$ are:
\begin{align}
&&K_{v_{c_\beta} v'_{c_\beta}}^{{\Lambda}(2)} &= \frac{1}{W_{d}} \int \int \Big[\chi_{v_{c_\beta}}^\Lambda(R) \MV^{(e)\Lambda}_{c_\beta,d}(R) F_{d} (R_<)\nonumber\\
&&\times &\:  G_{d}(R_>) \MV^{(e)\Lambda}_{d,c_\beta'}(R') \chi_{v'_{c_\beta}}^\Lambda (R') \Big]dR dR'
.
\end{align}
Here $W_{d}$ is the Wronskian between the regular ($F_{d}$) and irregular ({$G_{d}$) solutions at the origin of the nuclear Shr\"odinger equation, $R_<=min(R,R')$ and $R_>=max(R,R')$, while $\beta=1,2$.

\textit{iii) Computation of the eigenchannel wave functions:} It relies on the eigenvectors  and eigenvalues of the reaction matrix $\boldsymbol{\mathcal{K}}$, i.e. the columns of the matrix $\boldsymbol{{U}}$ and the elements of the diagonal matrix $\boldsymbol{\tan(\eta)}$ respectively:
\begin{align}\label{eq:EqnForU}
&&\boldsymbol{\mathcal{K}U}= -\frac{1}{\pi}\boldsymbol{\tan(\eta)U}
,
\end{align}
where the non-vanishing elements of the diagonal matrix $\boldsymbol\eta$ are the phase shifts introduced into the wave functions by the short-range interactions. 

\textit{iv) Frame transformation from the Born-Oppenheimer representation to the close-coupling one:} It is performed via the matrices $\boldsymbol{\mathcal{C}}$ and $\boldsymbol{\mathcal{S}}$, built on the basis of the matrices $\boldsymbol{{U}}$ and $\boldsymbol\eta$ and on the  quantum defect characterizing the incident/Rydberg electron, $\mu_{l}^{\Lambda}(R)$. The elements of these matrices are:
\begin{align}
\mathcal{C}_{v^+_{c_\beta},\Lambda \alpha} & =
     \sum_{v_{c_\beta}} U^{\Lambda}_{{v_{c_\beta}}, \alpha}
     \left\langle \chi_{v^+_{c_\beta}} (R)\left| \cos\left(\pi\mu^{\Lambda}_{c_\beta} (R) + \eta_{\alpha}^{{\Lambda}}\right) \right| \chi_{v_{c_\beta}}(R) \right\rangle, \label{C1}
\\
\mathcal{C}_{d,\Lambda\alpha} & = U^{\Lambda}_{d, \alpha} \cos\eta^{\Lambda}_\alpha, \label{C2}
\\
\mathcal{S}_{v^+_{c_\beta},\Lambda \alpha}  & =
     \sum_{v_{c_\beta}} U^{\Lambda}_{{v_{c_\beta}}, \alpha}
     \left\langle \chi_{v^+_{c_\beta}} (R)\left| \sin\left(\pi\mu^{\Lambda}_{c_\beta} (R) + \eta_{\alpha}^{{\Lambda}}\right) \right|\chi_{v_{c_\beta}}(R) \right\rangle, \label{S1}
\\
\mathcal{S}_{d,\Lambda\alpha}  & = U^{\Lambda}_{d, \alpha}  \sin\eta^{\Lambda}_\alpha, \label{S2}
\end{align}
where $\alpha$ denotes the eigenchannels built through the diagonalization of the reaction matrix $\Kmat$, while $\beta=1,2$.

\textit{v) Construction of the generalized scattering matrix $\boldsymbol{\mathcal{X}}$}, eventually split in blocks associated with energetically open and closed (o and c respectively) channels:
\begin{align}
&&\boldsymbol{\mathcal{X}}=\frac{\boldsymbol{\mathcal{C}}+i\boldsymbol{\mathcal{S}}}{\boldsymbol{\mathcal{C}}-i\boldsymbol{\mathcal{S}}}
\qquad
\boldsymbol{\mathcal{X}}= \left(\begin{array}{cc} \boldsymbol{X_{oo}} & \boldsymbol{X_{oc}}\\
                   \boldsymbol{X_{co}} & \boldsymbol{X_{cc}} \end{array} \right).
\end{align}

\textit{vi) Construction of the physical scattering matrix $\boldsymbol{S}$}, whose elements link mutually the open channels exclusively, given \mbox{by \cite{seaton1983}:}
\begin{align}\label{eq:solve3}
&&\boldsymbol{S}=\boldsymbol{X_{oo}}-\boldsymbol{X_{oc}}\frac{1}{X_{cc}-\exp(-i2\pi\boldsymbol{ \nu})}\boldsymbol{X_{co}},
\end{align}
where the matrix $\exp(-i2\pi\boldsymbol{ \nu})$ is diagonal and relies on the effective quantum numbers $\nu_{v^{+}}$ associated to the vibrational thresholds of the closed channels.

\textit{vii) Computation of the cross-sections:}
Given the target cation on its level $v_i^+$, its impact with an electron of energy $\varepsilon$ results in DR according to the \mbox{formula:}
\begin{align}\label{eq:eqDR}
&&\sigma _{{\rm diss} \leftarrow v_{i}^{+}}=\sum_{\Lambda,\mathrm{sym}} \frac{\pi}{4\varepsilon}\rho^{\Lambda,{\rm sym}}\mid S_{d,v_{i}^{+}}\mid^2,
\end{align}
or results in VE following the \mbox{formula:}
\begin{align}\label{eq:eqVE}
&&\sigma _{{\rm v_{f}^{+}} \leftarrow v_{i}^{+}}=\sum_{\Lambda,\mathrm{sym}} \frac{\pi}{4\varepsilon}\rho^{\Lambda,{\rm sym}}\mid S_{v_{f}^{+},v_{i}^{+}}-\delta_{v_{f}^{+},v_{i}^{+}}\mid^2,
\end{align}
where $\rho^{\Lambda,{\rm sym}}$ stands for the ratio between the multiplicity of the involved electronic states of the neutral and that of the target ion.

\subsection{Molecular data}

\begin{figure}
\centering
\includegraphics[width=\linewidth]{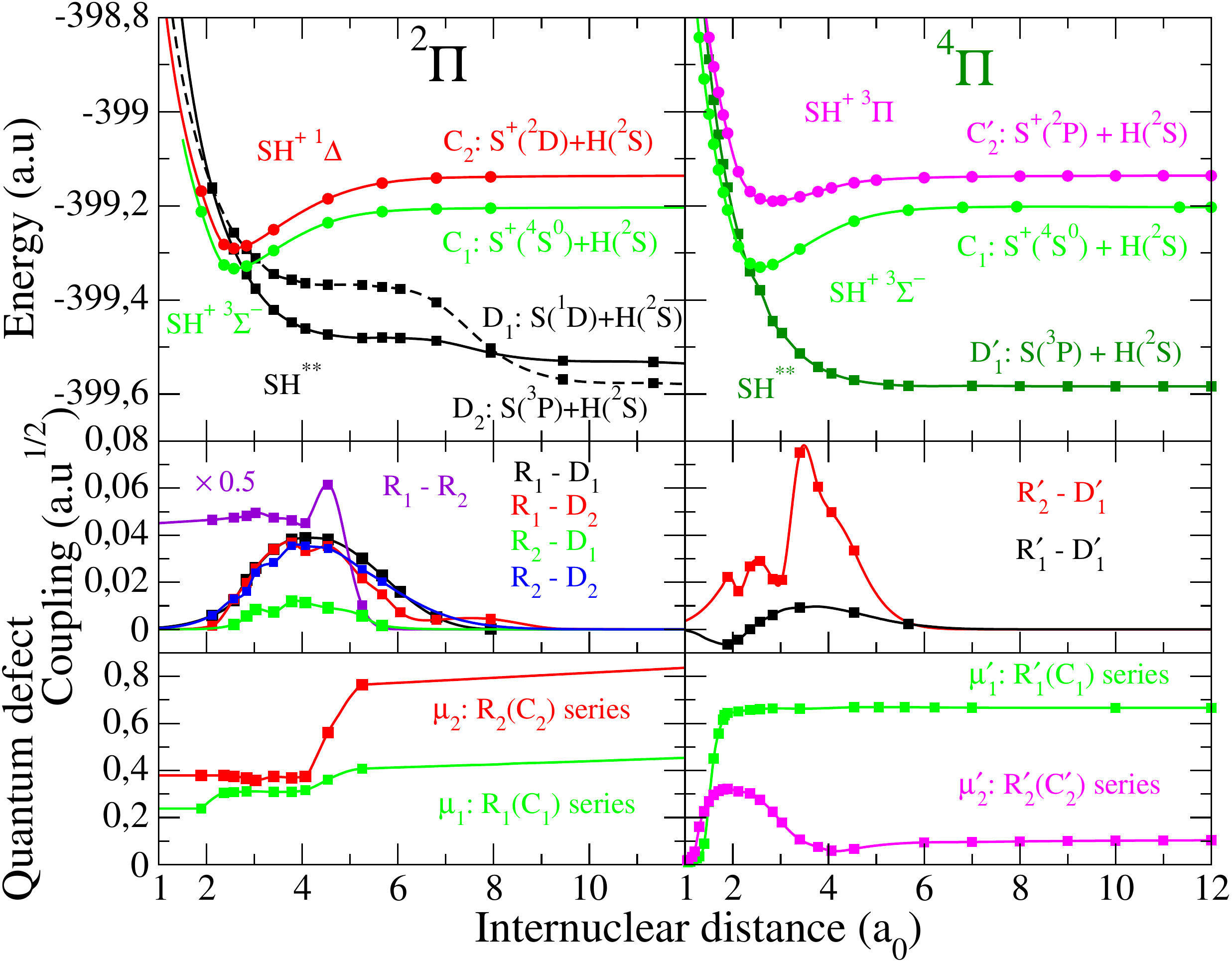}
\caption{Molecular structure data set: Left panels: $^2\Pi$ symmetry of SH~\cite{kashinski2017}, right panels: $^4\Pi$ symmetry of SH~\cite{kashinski2021}. First row: PECs; Second row: Rydberg-valence and Rydberg-Rydberg electronic couplings; Third row: Quantum defects. Symbols stand for the quantum chemistry data points while solid and dashed lines represent the interpolated and extrapolated data.
\label{fig:MolecularData}}
\end{figure}

The molecular data required by MQDT method is as follows: the PECs of the ground and excited ion states (C$_1$, C$_2$ and C$'_2$ in Fig.\ref{fig:MolecularData}), the PECs of the relevant dissociative autoionizing states of the neutral molecule (D$_1$, D$_2$ and D$'_1$), the quantum defects characterizing the different Rydberg series of bound predissociated states (R$_1$, R$_2$, R$'_1$ and R$'_2$), and the valence-Rydberg (R$_1$-D$_1$, etc.) and Rydberg-Rydberg  (R$_1$-R$_2$, etc.) electronic couplings between the dissociation and/or the ionization 
channels  of the neutral system.

The DR cross section (Eq. \ref{eq:eqDR}) is extremely sensitive to the position of the crossing point between the PECs of the neutral dissociative states with respect to the PECs of the target ion. A slight change in the position of the crossing point can lead to a significant change in the predicted cross section. Furthermore, the PECs of the dissociative states must also converge to the correct asymptotic limits for large values of the internuclear distance in order to be able to clearly define them as either energetically open or closed channels. Thus special care need to be done calculating these data sets.

Only a few theoretical methods are able to provide all the necessary molecular data with the desired accuracy. Among these is the R-matrix theory \cite{tennyson2010}, the complex Kohn variational method \cite{rescigno1995} and in the present case the block diagonalization method \cite{pacher1988} combined with GAMESS suite \cite{GAMESS}.

The molecular data sets relevant for our DR and VE calculation are summarized in Fig.~\ref{fig:MolecularData}: On the left panels we present those 
calculated before 
for the $^{2}\Pi$ symmetry of SH~\cite{kashinski2017}, while on the right panel those obtained recently for the $^{4}\Pi$ symmetry~\cite{kashinski2021}. The details for the $^{2}\Pi$ symmetry were given in~\cite{kashinski2017} and, consequently, here we are focusing on the $^{4}\Pi$ symmetry. 
In the first row we displayed the PECs for the ground ion core (SH$^+$ X $^3\Sigma^-$: C$_1$, green), for the excited ion core with $^3\Pi$ symmetry (C$^{'}_2$, magenta) and for the dissociative state of SH having $^4\Pi$ symmetry (D$^{'}_1$, dark green). In the second row we give the electronic couplings of the highly-excited Rydberg states (R$^{'}_1$ in black and R$^{'}_2$ in red) to the dissociative continuum (D$^{'}_1$). These coupling terms are  $\MV^{(e)\Lambda}_{d,c_\beta}(R)$ defined in Eqs.~(\ref{eq:elcoup}). The terms of $\MV^{(e)\Lambda}_{c_1,c_2}(R)$ in Eqs.~(\ref{eq:elcoupCores}) representing the coupling between the two Rydberg series as a result of the diabatization procedure are zero~\cite{kashinski2021}. In the third row we have the quantum defects: $\mu^{'}_{1}$ (green) and $\mu^{'}_{2}$ (magenta) that describe the infinite Rydberg series of PECs, converging to the ground core (C$_1$) and the excited core (C$^{'}_2$), respectively. The quantum defects were extracted from the Rydberg PECs, provided by the block diagonalization method.

\begin{table}
	\centering
	\caption{Fitting parameters D and C$_n$ used in the long-range multipole formula $\mbox{D}+\mbox{C}_n/R^n$ for the different molecular states.}
	\label{tab:1}
	\begin{tabular}{llcccc} 
		\hline
		\hline
		\multicolumn{2}{c}{Molecular} & state & n  & D & C$_n$\\
		 \multicolumn{2}{c}{system}    &           &    & (Hartree) & (Hartree$\times$Bohr$^n)$\\
		\hline
	SH$^+$    & $^3\Sigma^-$ & C$_1$ & 4& -399.20323325 & -16.769107 \\
	SH$^+$    & $^3\Pi$ & C$_2'$ & 4 & -399.13555828 & -6.073002 \\
	SH$^{**}$	& $^4\Pi$ & D$_1'$ & 6 & -399.58373325 & 89.132274 \\
	SH$^*$    & $^4\Pi$ & R$_1'$ & 6 & -399.29465171 & -232.669573 \\
	SH$^*$    & $^4\Pi$ & R$_2'$ & 6 & -399.27473902 & -146.713317 \\
		\hline
		\hline
	\end{tabular}
\end{table}

The original diabatic potential energy curves are given by full symbols in Fig.~\ref{fig:MolecularData}. The ion and neutral PECs have been extended towards large internuclear distances by adding a $\mbox{D}+\mbox{C}_n/R^n$ multipole term to the curves. The details for each molecular state are summarised in table~\ref{tab:1}. Moreover, in order to get the NIST atomic dissociation limits we have performed a global shift of $\Delta_{\text{NIST}}=1.45708$ Hartree for each of the PECs of the $^4\Pi$ symmetry set by preserving all other characteristics (e.g. ionization energies) of the electronic states. 

The quantum chemistry electronic couplings were fitted with gaussian functions chosen to match their peak values. This procedure provides smooth functions on both ends, by considering the behaviour of the calculated couplings at small internuclear distances (left end) and by assuring a steep but continuous fall as soon as the resonant dissociative state crosses the electronic ground state of the ion (right end).

\section{Results and discussion} \label{sec:results}

The cross sections quantifying the major features of the DR of SH$^+$  are displayed in Figures  \ref{fig:fig2} and \ref{fig:fig4}. Their corresponding thermal rate coefficients - Figures \ref{fig:fig5} and \ref{fig:fig7} - are derived by convoluting the cross sections with the Maxwell distribution function for velocities $v$ (related to incident energy of the electrons by $\varepsilon= \frac{1}{2} mv^{2} $) of the free electrons,
\begin{align}\label{eq:rate}
&&k(T)=\frac{8\pi}{\sqrt{m}(2\pi k_{B}T)^{3/2}} \int^{+\infty}_{0}\sigma(\varepsilon)\: \varepsilon \: exp(-\varepsilon/k_{B}T)\: d\varepsilon
,
\end{align}
\noindent
where $\sigma(\varepsilon)$ is given by (Eq. \ref{eq:eqDR}), $k_{B}$ is the Boltzmann constant and $T$ is the absolute temperature of the electrons. 

Figure \ref{fig:yields} gives the yields for the produced atomic fragments \mbox{S ($^3$P)+H ($^2$S)} and \mbox{S ($^1$D) +H ($^2$S)} via the DR process by considering the most relevant $^2\Pi$ and $^4\Pi$ symmetries of SH. 

Finally, Figures~\ref{fig:xsecve} and \ref{fig:rateve} present the vibrational excitation cross sections and rate coefficients respectively from the ground vibrational level of the ion to its two lowest excited vibrational levels, in comparison with those of the DR.

The calculations include $33$ ionization channels in total corresponding to $18$ vibrational levels belonging to the ground SH$^+$ X $^3\Sigma^-$ state (C$_1$ in Figure 1) and $15$ vibrational levels belonging to the excited SH$^+$ A $^3\Pi$ state (C$^{'}_2$ in Figure~\ref{fig:MolecularData}). For the incident electron we consider the $p$ partial wave only. The calculations are performed in second order of the K-matrix and include both \textit{direct} and \textit{indirect} mechanisms of the studied processes. The explored range for the incident electron energy is $1-500$ meV (with a step of $0.01$ meV) and the temperature range is $10-1000$ K.

\begin{figure}
\centering
\includegraphics[width=\linewidth]{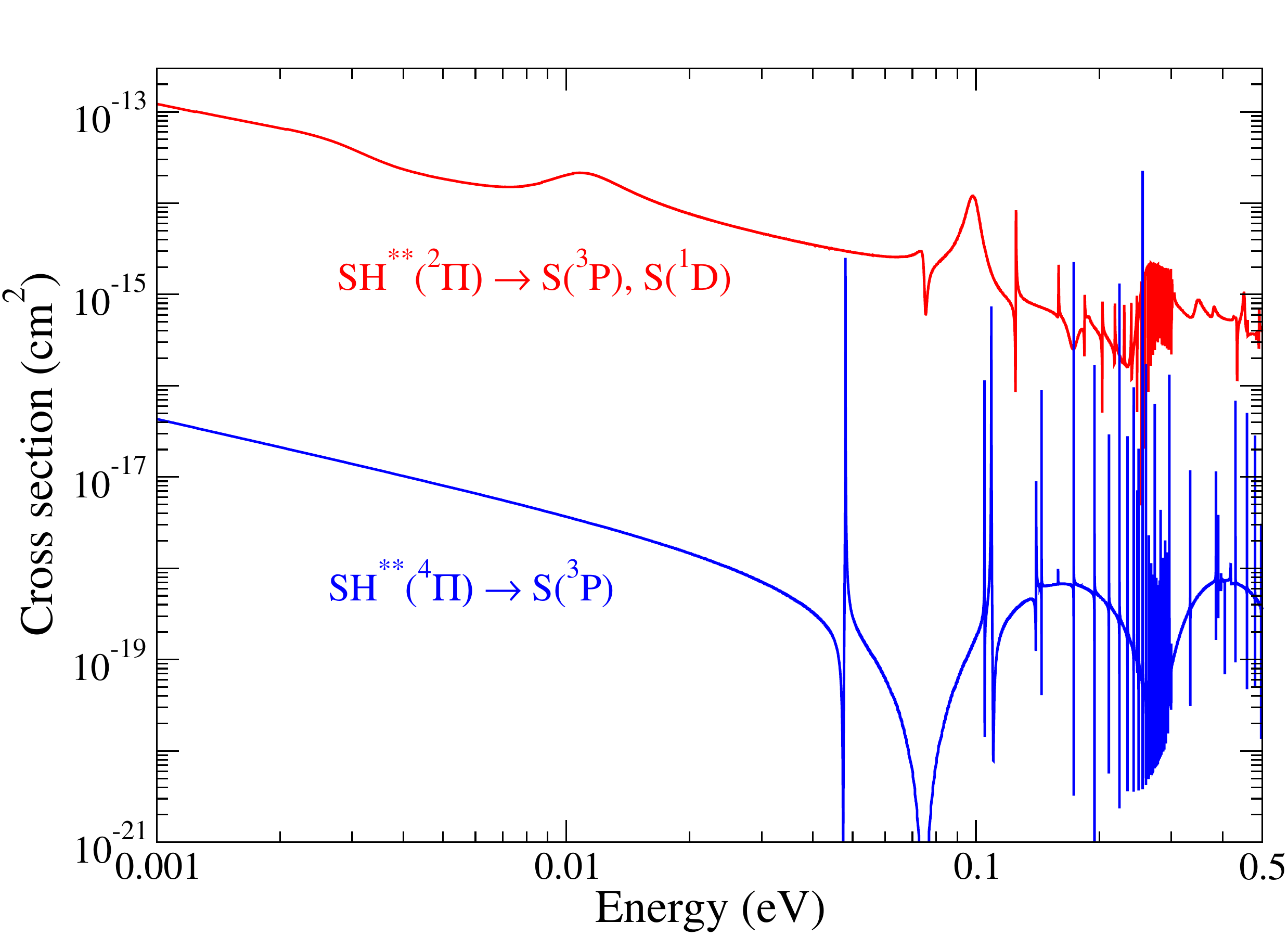}
\caption{Cross section for  SH$^+$ DR into the  state of $^4\Pi$ symmetry (D$^{'}_1$ in Figure~\ref{fig:MolecularData}, blue line) compared to that for DR into the states of  $^2\Pi$ symmetry (D$_1$ and D$_2$ in Figure~\ref{fig:MolecularData}, red line).
\label{fig:fig2}}
\end{figure}

\begin{figure}
\centering
\includegraphics[width=\linewidth]{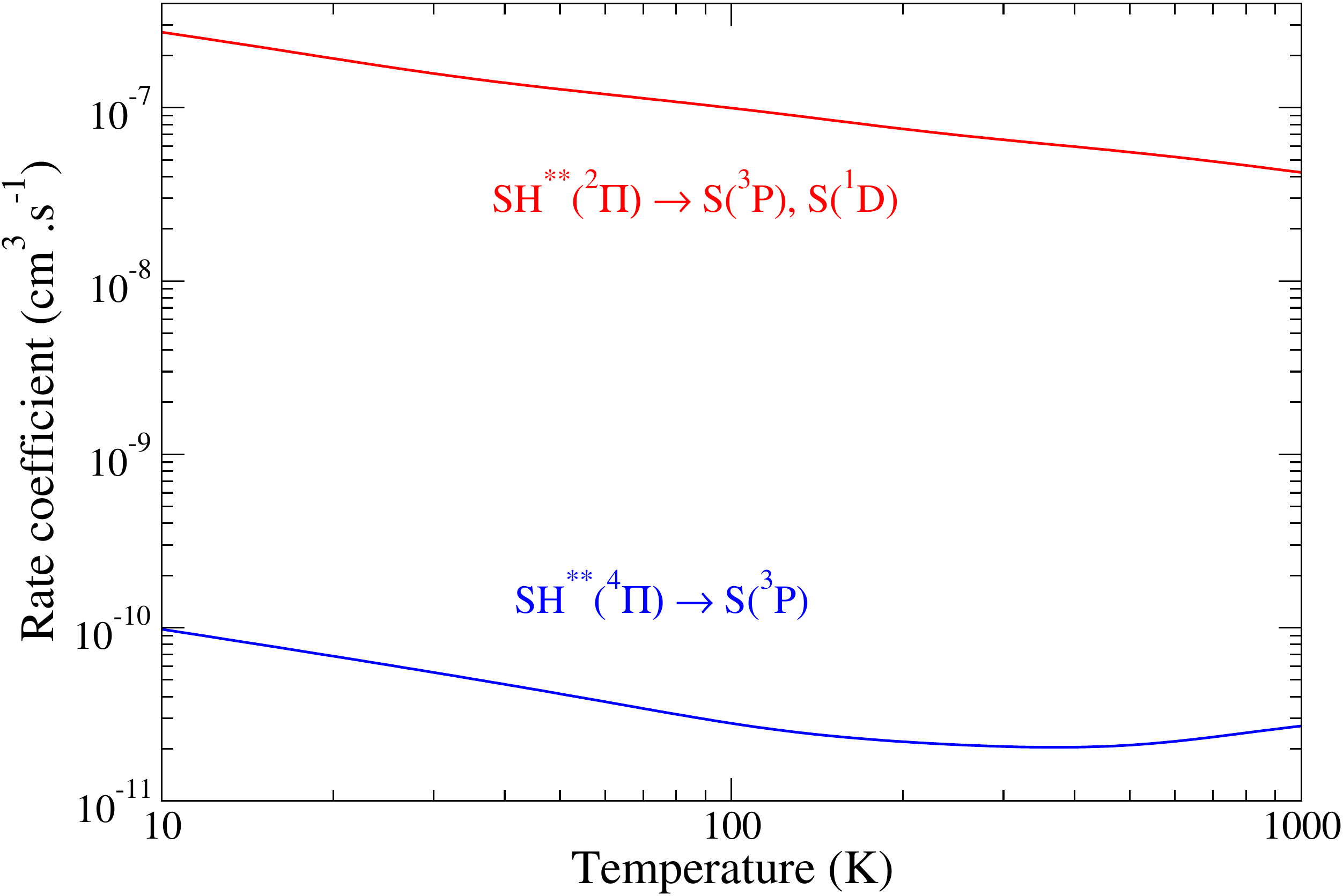}
\caption{Thermal rate coefficients calculated from the cross sections presented in  Fig. \ref{fig:fig2}.
\label{fig:fig5}}
\end{figure}

\subsection{Dissociative recombination}

The calculated total (\textit{direct} and \textit{indirect} mechanisms) DR cross section for the $^4\Pi$ symmetry  - blue curve in Figure \ref{fig:fig2} - is presented and compared with the previously obtained cross sections for the $^2\Pi$ symmetry - red curve on the same figure,  computed in our previous study~\cite{kashinski2017}. 
One can clearly observe the $1/energy$ background behaviour of the DR cross sections at low collision energies for both molecular symmetries. Another typical characteristic of the total cross sections is the presence of resonance structures. They originate from the temporary capture of the incident electron into highly excited Rydberg states 
built on the ionic cores  of PEC's C$_1$ and C$_2$ for the  $^2\Pi$ symmetry and C'$_1$ for the $^4\Pi$ one (see Figure~\ref{fig:MolecularData}).

One may notice that at low collision energies the contribution of the $^4\Pi$ symmetry is between three and four orders of magnitude smaller than that of the $^2\Pi$ symmetry, as illustrated in Fig. \ref{fig:fig5}. Indeed, the $^2\Pi$ symmetry contributes with a rate around $10^{-7}$ cm$^3\cdot$s$^{-1}$, while the $^4\Pi$ symmetry provides around a few $10^{-11}$ cm$^3\cdot$s$^{-1}$ at 100 K. This can be now - after performing our calculations - explained by the less favourable crossing of the dissociative state PEC with that of the target ion one - top right-hand side part of Fig.~\ref{fig:MolecularData} - and the relatively small valence-Rydberg electronic couplings in the Franck-Condon region, middle right-hand-side part of the same figure.

One can conclude that the $^4\Pi$ symmetry states of SH are far less important than the $^2\Pi$ ones, contrary to our expectation~\cite{kashinski2017}, and they do not contribute significantly to a better agreement of our theoretical results with the storage-ring experiment at very low energy.

\begin{figure}
\centering
\includegraphics[width=\linewidth]{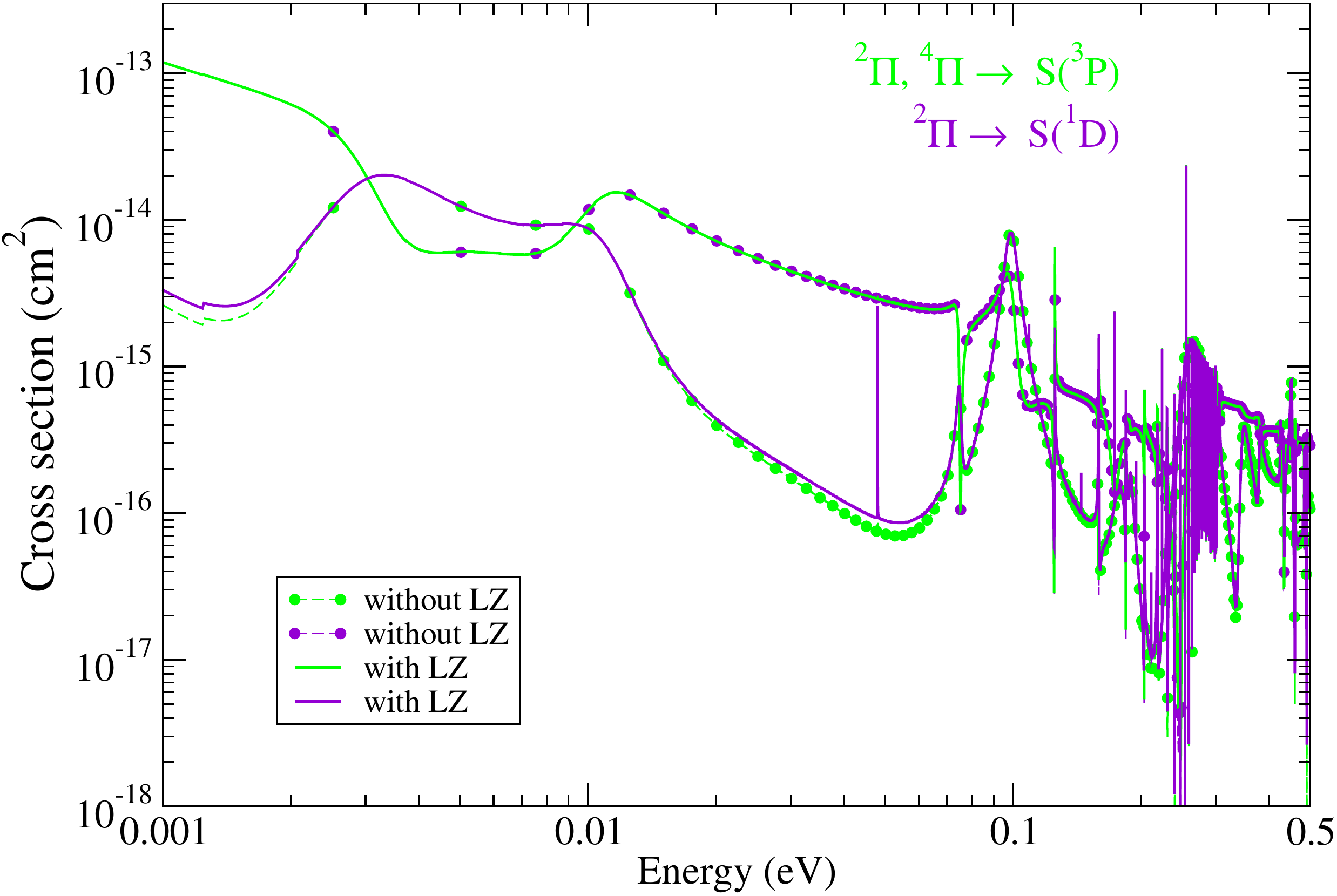}
\caption{
DR cross sections contributing to the formation of \mbox{S($^3$P)+H($^2$S)} (green) and \mbox{S($^1$D)+H($^2$S)} (violet) atomic fragments. The dashed lines with circles stand for the results without the Landau-Zener (LZ) treatment, whereas the continuous lines stand for the final results, obtained following the Landau-Zener treatment.
\label{fig:fig4}}
\end{figure}
\begin{figure}
\centering
\includegraphics[width=\linewidth]{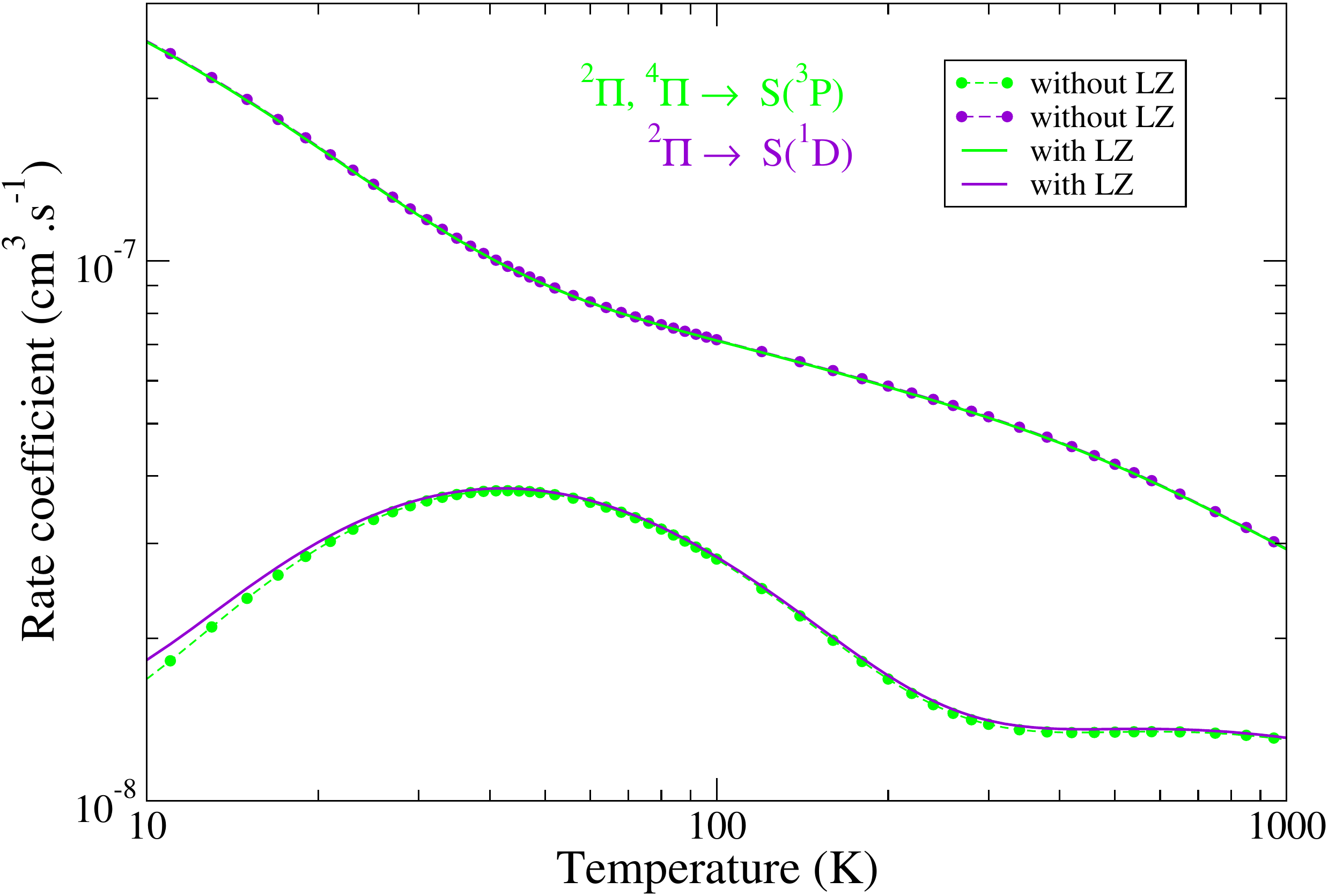}
\caption{
Thermal rate coefficients of the production of ground and excited sulfur atoms via DR, calculated from the cross sections presented in Fig. \ref{fig:fig4}. The legend is the same as that in this figure.
\label{fig:fig7}}
\end{figure}

\subsection{Branching ratios}

In order to  understand the Sulfur chemistry, besides the cross sections and thermal rate coefficients of DR, an important issue is the final state distribution of the \mbox{S} atoms resulting from this process. 

\begin{table*}
	\centering
	\caption{
	Landau-Zener probabilities (eq.~(\ref{eq:LZprob})) for the  transition  between the two lowest dissociative $^2\Pi$ diabatic states of the SH molecule - D$_1$ and D$_2$  in the upper-left panel of fig.~\ref{fig:MolecularData} - and corresponding branching ratios as function of the collision energy. Our MQDT/Landau-Zener results are compared with the TSR measured data, when available ~\cite{becker2015,becker2016}.}
	\label{tab:tab2}
	\begin{tabular}{ccccccccc} 
		\hline\hline
& & \multicolumn{2}{c} {MQDT--no LZ} &\multicolumn{2}{c} {MQDT--LZ} & \multicolumn{3}{c} {TSR} \\[1pt]
Energy	& $P_{12}	$  & $n(^3$P) & $n(^1$D)& $n(^3$P) & $n(^1$D) & $n(^3$P) & $n(^1$D) & $n(^1$S)\\
(eV)	&  &  &  & &  & \\
		\hline\hline
0.001& 0.99403 & 0.274 & 0.726 & 0.724 & 0.276 & 0.61 & 0.33 & 0.06\\
0.01	&  0.99400 & 0.140 & 0.860 & 0.856 & 0.144 & ---    & ---& ---	\\
0.1	& 0.99373	 & 0.616 &	 0.384 & 0.386 & 0.614 & ---    & ---& ---	\\
0.5	& 0.99246	& 0.320 &	0.680 & 0.678 & 0.322 & ---    & ---& ---	\\
		\hline\hline
	\end{tabular}
\end{table*}

According to Fig.~\ref{fig:MolecularData}, and neglecting - in a first step - the possibility of re-distribution of the probability flux at the crossing between the PECs of the two dissociative paths,  \mbox{S} atoms in their ground state $^3$P are produced via the capture into  D$_2$ $^2\Pi$ and D$^{'}_1$  $^4\Pi$ states, whereas the excited state $^1$D is populated via  the D$_1$  $^2\Pi$ dissociative path.

Figs. \ref{fig:fig4} and \ref{fig:fig7} illustrate these branching ratios. One can notice that in the range  $0.001-0.5$ eV the cross section of generation of \mbox{S($^3$P)} atoms - green dashed lines with green circles - is  overall smaller than the one obtained for \mbox{S($^1$D)} - violet dashed lines with violet circles - up to two orders of magnitude. As an exception, around $0.005$ eV and around $0.09$ eV collision energies, the DR cross section leading to \mbox{S($^3$P)} is higher than that of \mbox{S($^1$D)} by about one order of magnitude. 
The corresponding thermal rate coefficients are presented in figure \ref{fig:fig7}, with the same color and symbol code. The generation of \mbox{S($^3$P)} atoms (green dashed line with green circles) has an average rate of $2\cdot10^{-8}$ cm$^3$.s$^{-1}$ in the temperature range $10-200$ K. It reaches its peak value of $4\cdot10^{-8}$ cm$^3$.s$^{-1}$ at about $40$ K, having its origin in the behaviour of the cross section in the energy range  $4-10$ meV. This rate is practically constant above 250 K. On the other hand, the rate of production of \mbox{S($^1$D)} atoms (violet line with violet circles) monotonically decrease from $2\cdot10^{-7}$ cm$^3$.s$^{-1}$ 
at $10$ K to $3\cdot10^{-8}$ cm$^3$.s$^{-1}$  at $1000$ K. Obviously, in this first-step approach, the production of \mbox{S($^1$D)} atoms dominates largely that of \mbox{S($^3$P)} atoms at low temperature.

\begin{figure}
\centering
\includegraphics[width=\linewidth]{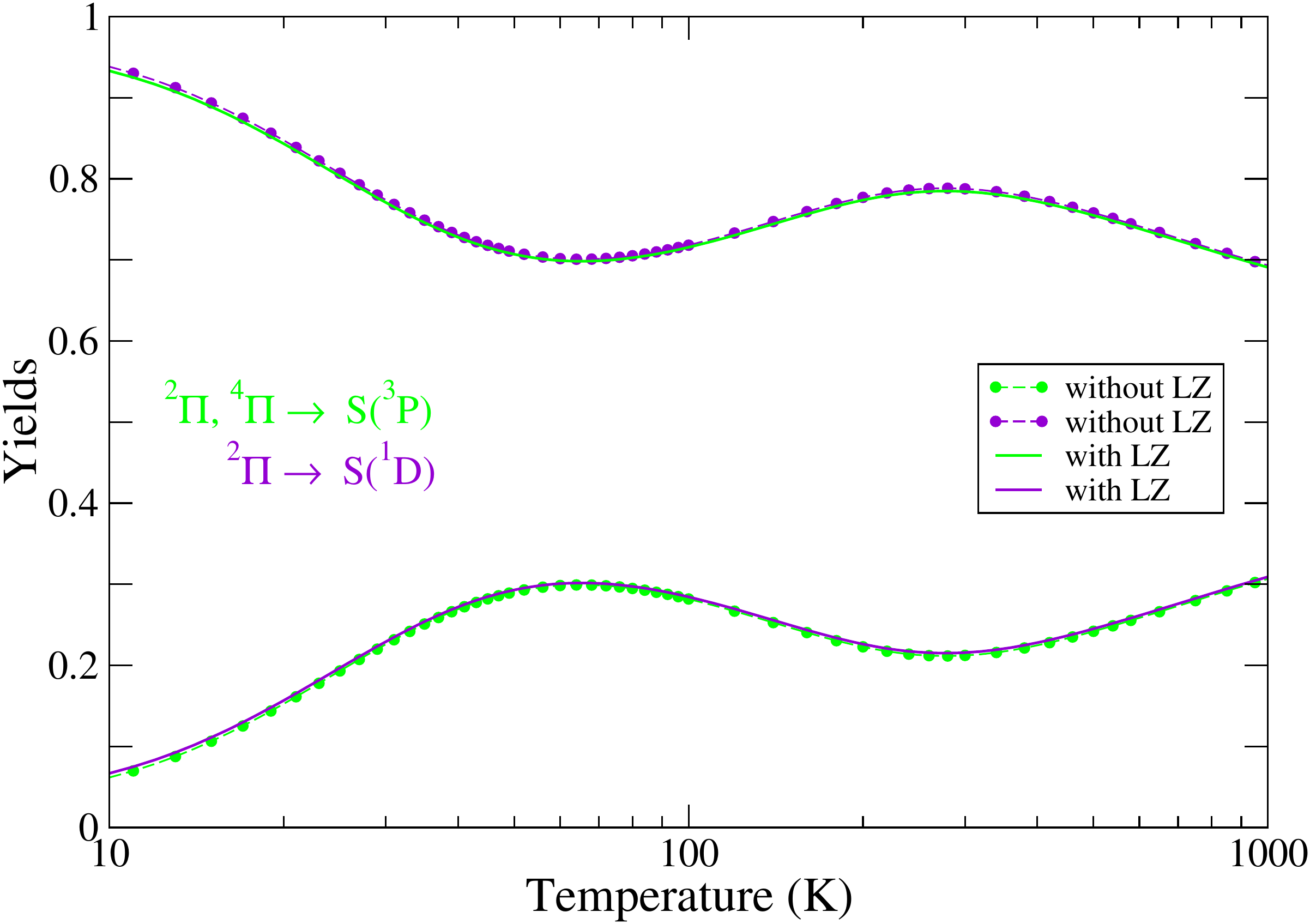}
\caption{Yields for the production of ground and excited sulfur atoms via DR, calculated from the rate coefficients of Fig. \ref{fig:fig7} .The legend is the same as that in Fig. 4.
\label{fig:yields}}
\end{figure}

To better compare the processes of populating the two lowest energy levels of the S atom, we present in Fig. \ref{fig:yields} the yields resulting from the branching ratios from Fig.  \ref{fig:fig7}, calculated by dividing the corresponding thermal rate coefficients to the sum of both. 
In average, in the electron temperature range 10-1000 K, the DR process contributes to the generations of \mbox{S($^3$P)} atoms as much as $24\%$, and to that of \mbox{S($^1$D) }atoms with  $76\%$. 

In order to account for the flux redistribution at the crossing point $(R_0\sim 8\,\,a_0)$ between the PECs of the D$_1$ and D$_2$ dissociative states, we performed a Landau-Zener calculation~\cite{zener1932,rubbmark1981}. 
The probability of  transition from one diabatic branch  to the other - $P_{12}$ - is given by~\cite{asa2000}
\begin{equation}\label{eq:LZprob}
P_{12}(\varepsilon)=1-e^{-\frac{2\pi V^2_{12}(R_0)}{\hbar av}},
\end{equation}
where $V_{12}(R_0)=0.0314$ Hartree is the coupling element between the two dissociative states at their crossing point~\cite{kashinski2017}, $a=\frac{dV_1}{dR}-\frac{dV_2}{dR}$ stands for the differences of PEC slopes at the crossing point and $v$ is the relative collisional velocity~\cite{asa2000}.  

Consequently, starting from the dissociative states D$_1$ and D$_2$ (cf. the upper-left panel of Fig.~\ref{fig:MolecularData}), one obtain the cross sections $\sigma'_1(\varepsilon)$ and $\sigma'_2(\varepsilon)$ towards the S($^3$P)$+$H($^2$S) and S($^1$D)$+$H($^2$S) dissociation limits, respectively, according to the formulas:
\begin{eqnarray}\label{eq:lz}
\sigma'_1(\varepsilon)&=&(1-P_{12})\sigma_1(\varepsilon) + P_{12}\sigma_2(\varepsilon)\nonumber\\
\sigma'_2(\varepsilon)&=&(1-P_{12})\sigma_2(\varepsilon) + P_{12}\sigma_1(\varepsilon).
\end{eqnarray}

Table~\ref{tab:tab2} contains these Landau-Zener transition probabilities $P_{12}$ as function of the collision energy.
Moreover, we display the branching ratios resulting from the cross sections convoluted with the anisotropic Maxwell distribution of the relative electron$/$ion velocities corresponding to longitudinal and transversal temperatures $T_{\parallel}= 0.025$ meV and $T_{\perp}= 1.65$ meV respectively, before and after performing the Landau-Zener calculation. This convolution allowed us to compare our branching ratios with those measured in the TSR storage ring for 1 meV collision energy ~\cite{becker2015,becker2016}.

The re-distribution of the flux of probability at the crossing between D$_1$ and D$_2$ results in the cross sections and yields illustrated in thick lines in Figs.~\ref{fig:fig4},~\ref{fig:fig7} and~\ref{fig:yields}, and in the yields displayed in Table~\ref{tab:tab2}. One may notice in these figures an interchange of the populations in the ground and lowest excited state of the Sulfur atoms, due to this re-distribution. The good agreement with the TSR-measurements at very low energy supports this interchange, whereas the theoretically-computed yields at 1 meV without the Landau-Zener treatment disagree with the experimental ones.

\subsection{Vibrational excitation}

In addition to the DR cross sections and rate coefficients we have calculated vibrational excitations (VE) of the relaxed $(v_i^+=0)$ molecular target into the lowest excited vibrational states $(v_i^+= 1, 2)$ via the capture into states of both  $^2\Pi$ and $^4\Pi$ symmetries of the neutral. 
 
 The cross sections over symmetries and the global thermal rate coefficients summed over all symmetries, can be seen in Figs.~\ref{fig:xsecve} and \ref{fig:rateve} respectively, together with those obtained for DR. 
 
According to Fig.~\ref{fig:xsecve} the VE cross sections show sharp thresholds corresponding to the excitation energies of the ion. Similarly to DR, they are driven by the Rydberg resonances and for this process too we found smaller cross sections for the $^4\Pi$ molecular symmetry. However the magnitude of the obtained difference for the two symmetries is different, while for the DR we get a difference of about two orders of magnitude, the VE cross sections differ only by one order of magnitude.

\begin{figure}
\centering
\includegraphics[width=\linewidth]{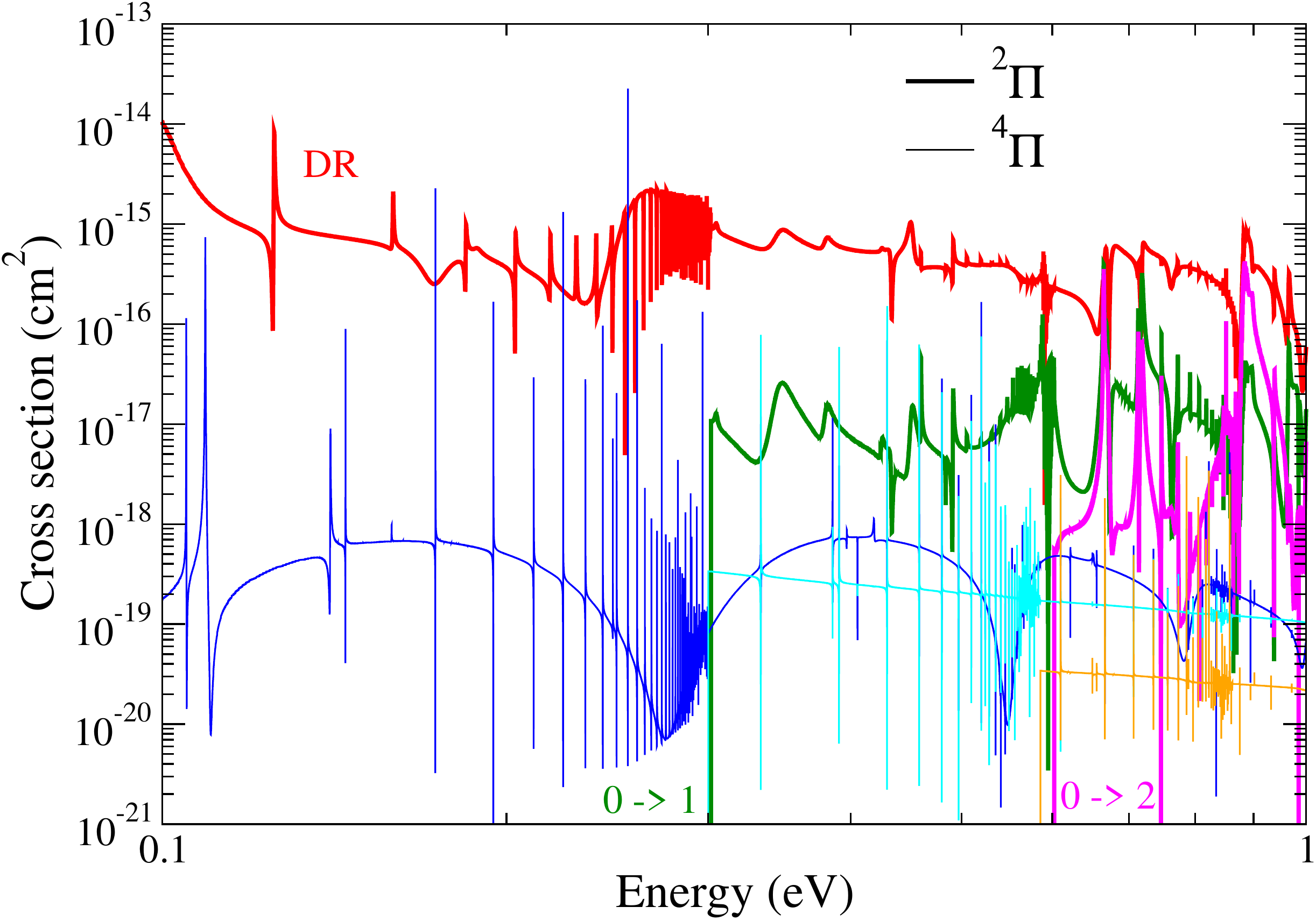}
\caption{Vibrational excitation and dissociative recombination cross sections of the vibrationally relaxed SH$^+$ molecular cation by considering both $^2\Pi$ and $^4\Pi$ symmetries of the formed molecular complex.
The red and blue curves stand for DR, the dark green and cyan ones for $0\rightarrow1$ VE and those in turquoise and maroon give VE for $0\rightarrow2$. Finally, the thick curves correspond for the $^2\Pi$ while the thin curves for the $^4\Pi$ molecular symmetry of the neutral. 
\label{fig:xsecve}}
\end{figure}

\begin{figure}
\centering
\includegraphics[width=\linewidth]{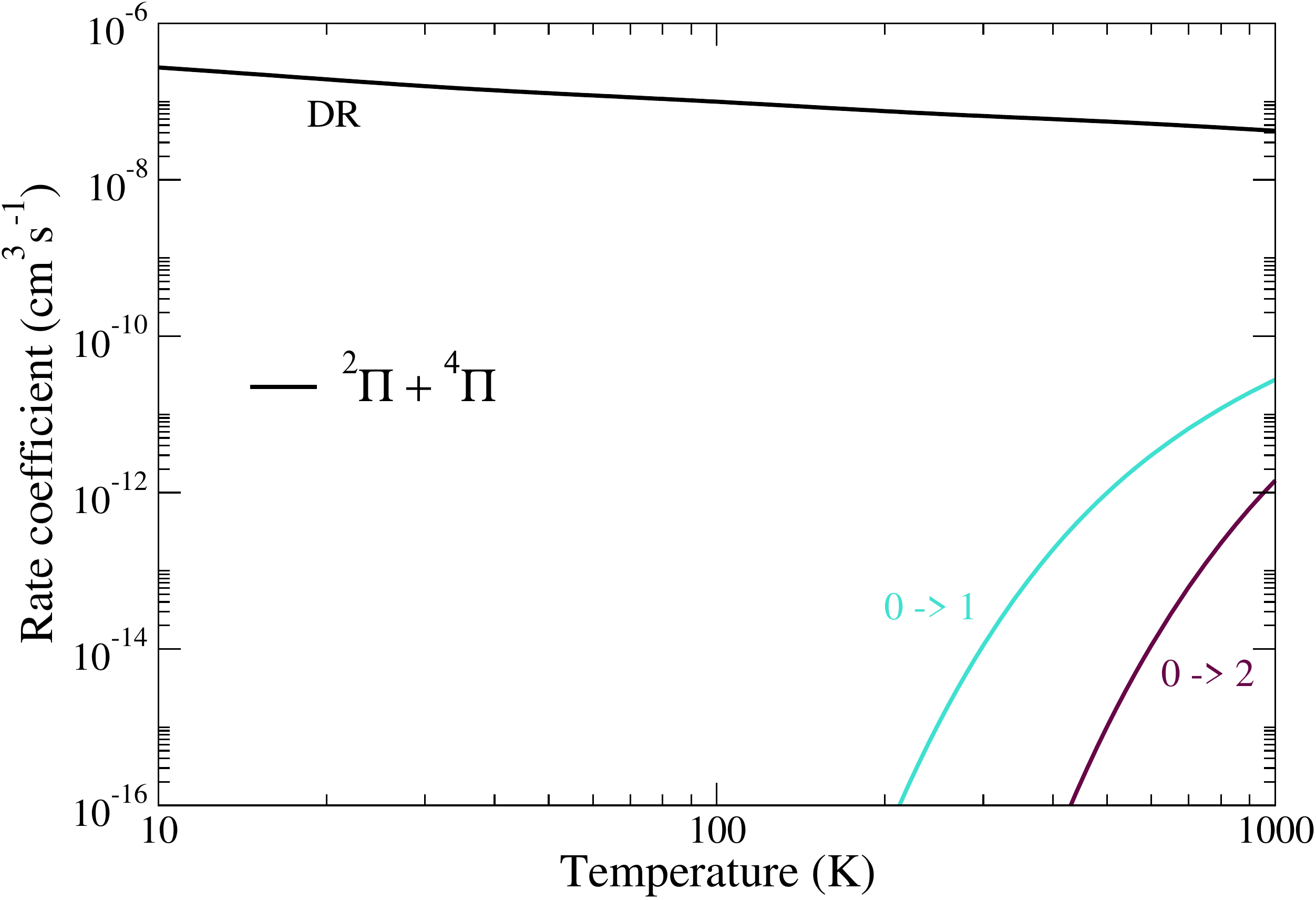}
\caption{
Vibrational excitation and dissociative recombination rate coefficients of the vibrationally relaxed SH$^+$ molecular cation:  sum of the contributions via the SH states of $^2\Pi$ and $^4\Pi$ symmetries.
\label{fig:rateve}
}
\end{figure}

\begin{table*}
	\centering
	\caption{Fitting parameters for the formula (\ref{eq:fit-RE}),  corresponding to the rate coefficients summed over $^2\Pi$ and $^4\Pi$ symmetries for dissociative recombination and vibrational excitation 
	of vibrationally relaxed ($v_{i}^{+}=0$) molecular cation. The fit applies for temperatures for which the rate coefficient is higher than $10^{-16}$ cm$^3$s$^{-1}$.
	}
	\label{tab:rate_fitt}
	\begin{tabular}{lccccccc} 
		\hline\hline
	&	 $v_{i}^{+}$  & $v_{f}^{+}$ & Temperature range & a$_0$ & a$_1$ & a$_2$ & RMS \\
	&  &  & (K) & (cm$^{3}$ s$^{-1}$K$^{-A{_1}}$) &  & (K) & \\
		\hline\hline
DR &  0	&  &	$10 \le T \le1000$ & 6.53104$\times10^{-8}$ & -0.348405 	& -2.59354	& 0.00855	\\
 \hline
VE &  0	& 1 & $220\le T\le400$ & 1.46041$\times10^{-9}$ & -0.432013 	& 3535.88 	& 0.00273	\\
 &   & &	$400 < T \le1000$ & 3.83266$\times10^{-10}$ & 0.364274 	& 3063.89	& 0.02488	\\
\hline
 VE &   0	& 2 &	$440 \le T \le1000$ & 8.44578$\times10^{-10}$ & 0.464803 & 6941.58 	& 0.01119	\\
		\hline\hline
	\end{tabular}
\end{table*}

Figure~\ref{fig:rateve} shows that while the DR rate coefficient is smoothly decreasing as a function of the temperature, the VE ones have a monotonically increasing behaviour due to the sharp thresholds. One can clearly see that the DR dominates over VE for all electronic temperatures up to 1000 K.

In order to facilitate the use of our recombination and excitation rate coefficients for kinetic modelling, we have fitted their temperature dependence by using the 
Arrhenius-type formula:
\begin{equation}
k(T) =a_0\left(\frac{T}{300}\right)^{a_1}e^{-\frac{a_2}{T}}, 
\label{eq:fit-RE}
\end{equation}
\noindent 
where the temperature $T$ is in Kelvin and the rate coefficient $k$ in cm$^{3}$s$^{-1}$. The fitting parameters for the DR of the ground vibrational state of the target and for the  VE into the lowest two vibrationally-excited states are summarized in Table~\ref{tab:rate_fitt}. For all the processes, the fitted values reproduce well our MQDT rate coefficients, according to the RMS values given in the forth column of each table in the whole temperature range from $10$ to $1000$ K.  

\section{Conclusions} \label{sec:concl}

In this paper we studied by MQDT the relevance of the $^4\Pi$ symmetry states of SH for the dissociative recombination of SH$^+$ at low energy/temperature. We have found that this symmetry has a far smaller contribution to the cross sections/rates than the $^2\Pi$ ones. Therefore, the comparison between theory and experiment  did not change after the present study: an overall good agreement from 10 meV to 1 eV, and a disagreement increasing progressively up to one order of magnitude when the energy is decreased from 10 to 0.1 meV. 

We produced the branching ratios, and we took into account the redistribution of the probability fluxes at the crossing between the PECs of the two dissociative states.
A spectacular inversion in the branching ratios takes place, resulting in a satisfactory agreement between our theoretical MQDT-Landau-Zener final results   and those produced by the storage ring TSR at very low energy. More than two third of the recombination events result in ground state sulfur atoms.

We also computed vibrational excitation cross sections and rate coefficients. This process is largely dominated by the dissociative recombination below 1000 K.

Further studies will be devoted to the account of the rotational effects, which may result in a better agreement between theory and experiment, especially in the very low energy range.

\section*{DATA AVAILABILITY}
The data underlying this article will be shared on reasonable request to the corresponding authors.

\section*{Acknowledgements}

The authors acknowledge support from La R\'egion Normandie, FEDER, and LabEx EMC3 via the projects PTOLEMEE, Bioengine COMUE Normandie Universit\'e, the Institute for Energy, Propulsion and Environment (FR-IEPE), the European Union via COST (European Cooperation in Science and Technology) actions TUMIEE (CA17126), MW-Gaia (CA18104) and MD-GAS (CA18212).
The authors are indebted to Agence Nationale de la Recherche (ANR) via the project MONA. This work was supported by the Programme National “Physique et Chimie du Milieu Interstellaire” (PCMI) of CNRS/INSU with INC/INP co-funded by CEA and CNES. J.Zs.M. thanks the financial support of the National Research, Development and Innovation Fund of Hungary, under the K 18 and FK 19 funding schemes with projects no. K 128621 and FK 132989, and it is very grateful for the hospitality of the Faculty of Physics of West University of Timisoara during his Visiting Professor Fellowship.


%
%





\end{document}